# All-optical image classification through unknown random diffusers using a single-pixel diffractive network

*Yi Luo*[1,2,3,+]          e-mail: yluo2016@ucla.edu
*Bijie Bai*[1,2,3,+]       e-mail: baibijie@g.ucla.edu
*Yuhang Li*[1,2,3]         e-mail: yuhangli@g.ucla.edu
*Ege Çetintaş*[1,2,3]      e-mail: egecetintas1@g.ucla.edu
*Aydogan Ozcan*[1,2,3*]    e-mail: ozcan@ucla.edu

[1]Electrical and Computer Engineering Department, University of California, Los Angeles, California 90095, USA

[2]Bioengineering Department, University of California, Los Angeles, California 90095, USA

[3]California Nano Systems Institute (CNSI), University of California, Los Angeles, California 90095, USA

[+]Equal contributing authors

[*]Correspondence: Prof. Aydogan Ozcan    E-mail: ozcan@ucla.edu

# Abstract

Classification of an object behind a random and unknown scattering medium sets a challenging task for computational imaging and machine vision fields. Recent deep learning-based approaches demonstrated the classification of objects using diffuser-distorted patterns collected by an image sensor. These methods demand relatively large-scale computing using deep neural networks running on digital computers. Here, we present an all-optical processor to directly classify unknown objects through unknown, random phase diffusers using broadband illumination detected with a single pixel. A set of transmissive diffractive layers, optimized using deep learning, forms a physical network that all-optically maps the spatial information of an input object behind a random diffuser into the power spectrum of the output light detected through a single pixel at the output plane of the diffractive network. We numerically demonstrated the accuracy of this framework using broadband radiation to classify unknown handwritten digits through random new diffusers, never used during the training phase, and achieved a blind testing accuracy of 88.53%. This single-pixel all-optical object classification system through random diffusers is based on passive diffractive layers that process broadband input light and can operate at any part of the electromagnetic spectrum by simply scaling the diffractive features proportional to the wavelength range of interest. These results have various potential applications in, e.g., biomedical imaging, security, robotics, and autonomous driving.

# Introduction

Imaging and recognizing objects through scattering media have been challenging in many fields, including biomedical imaging[1,2], oceanography[3,4], security[5], robotics[6], and autonomous driving[7,8], among others[9]. Numerous computational solutions have been developed to reconstruct an image distorted by a diffuser: deconvolution algorithms were used when the transmission matrix of a diffuser can be pre-measured as prior information[10]; adaptive optics and wavefront shaping methods were used with the help of guide-stars or reference objects[11,12]; iterative algorithms were used to solve for the images of the hidden objects utilizing the memory effect of a diffuser[13,14]. Multispectral or time-gated imaging methods were also used to bring additional degrees of freedom to recover the hidden objects[15,16], and similarly, deep neural networks were used to learn the features of diffusers and generalize to see through them[17,18]. Despite their success, for each object to be imaged, all these methods require access to large-scale computing provided by digital computers, which also hinders the practical frame rate of these computational imaging modalities. Furthermore, additional energy is consumed on the downstream tasks such as object recognition and image classification. Partially motivated by the fact that merging the two steps (image reconstruction and classification) could potentially reduce the energy consumption and computing time, deep learning-based digital solutions to directly classify objects hidden behind scattering media have also been demonstrated[19–21], which predicted the object class using speckle patterns as inputs without any digital image reconstruction. As a limitation, these deep learning-based methods can classify different objects through only the diffusers used in their training process, lacking generalization to blind, new diffusers that were never used in the training phase.

Recent works have presented an all-optical method to image through unknown diffusers using diffractive deep neural networks (D$^2$NNs), enabling passive, computer-free image reconstruction at the speed of light propagation through thin optical layers[22,23]. Diffractive networks form an all-optical machine learning platform that computes a given task using light diffraction through successive transmissive layers[24]. Each diffractive layer typically consists of tens of thousands of diffractive units (termed 'diffractive neurons') that modulate the phase and/or amplitude of the incident light. Deep learning tools, such as error backpropagation, are used to optimize the modulation values (e.g., transmission coefficients) of each layer, mapping a complex-valued input field containing the optical information of interest (to-be-processed) onto a desired output field. Computing using diffractive networks possesses the benefits of high speed, parallelism and low power consumption: the computational task of interest is completed while the incident light passes through passive thin diffractive layers at the speed of light, requiring no energy other than the illumination. This framework's success and capabilities were demonstrated numerically and experimentally by achieving various computational tasks, including object classification[24–27], hologram reconstruction[28], quantitative phase imaging[29], privacy-preserving class-specific imaging[30], logic operations[31,32], universal linear transformations[33], polarization processing[34] among others[35–44]. Diffractive networks can also process and shape the phase and amplitude of broadband input spectra to perform various tasks such as pulse shaping[45], wavelength-division multiplexing[46] and single-pixel image classification[47].

Here, we demonstrate broadband diffractive networks to directly classify unknown objects (e.g., MNIST handwritten digits[48]) through unknown, random diffusers using a single-pixel spectral detector (Fig. 1(a)). This broadband diffractive architecture uses 20 discrete wavelengths, mapping a diffuser-distorted complex optical field containing the spatial information of an input object into a spectral signature detected through a single pixel. A differential detection scheme was applied to the single-pixel output spectrum by assigning the intensities of 10 pre-determined wavelengths as the positive scores and the intensities of the remaining 10 pre-determined wavelengths as the negative scores, revealing the *differential spectral class scores* used for image classification through a single-pixel (Fig. 1(b)). During each training epoch, *n* different random phase diffusers with the same correlation length were used to generate unique distortions to the input objects. A loss function that penalized the classification accuracy through these random diffusers was used to optimize the modulation values on each diffractive layer. After being trained for 100 epochs, the single-pixel diffractive network successfully generalized to directly classify unknown handwritten digits completely hidden by unknown random phase diffusers never seen before during the training. After this one-time training, the resulting diffractive layers can be physically fabricated to form a passive single-pixel network that computes the desired classification task using only the illumination light, without a digital computer. In our numerical simulations, this single-pixel broadband diffractive network achieved a blind testing accuracy of 88.53%, successfully classifying handwritten digits through 80 randomly selected unknown phase diffusers, each with a correlation length of $25\lambda_{max}$, where $\lambda_{max}$ is the longest wavelength used in the illumination spectrum. These single-pixel diffractive designs that generalized to classify objects through unknown random diffusers can operate at any part of the electromagnetic spectrum (without the need for redesigning or retraining) by simply scaling the size of the diffractive features with respect to the operational wavelength range of interest.

Single-pixel all-optical diffractive image classification through random new diffusers presents a time- and energy-efficient solution for sensing and image classification through scattering media, with numerous potential applications in different fields, such as surveillance cameras, biomedical imaging and autonomous driving.

## Results

**Design of a broadband single-pixel diffractive network to classify handwritten digits through random unknown diffusers**

Broadband single-pixel diffractive networks were designed to classify MNIST handwritten digits placed behind unknown random diffusers using three successive diffractive layers and 20 discrete illumination wavelengths uniformly selected between a $\lambda_{min}$ and a $\lambda_{max}$. The spatial information of each handwritten digit placed at the input plane was encoded into the amplitude channel of all the 20 wavelengths. A random phase diffuser with a correlation length of $25\lambda_{max}$ was placed $33\lambda_{max}$ away from the input object along the optical axis to create random distortions to the optical field (Fig. 2(a)). The distances from the random diffuser to the first diffractive layer and between two successive diffractive layers were set to be $25\lambda_{max}$. The level of image distortion due to a random diffuser (with a correlation length of $25\lambda_{max}$) is visualized in Fig. 1(c) for $\lambda_{min}$ and $\lambda_{max}$, separately, as well as for all the 20 illumination wavelengths simultaneously on, shown for comparison.

Each distorted optical field at a given wavelength was forward propagated through three successive diffractive layers, each composed of 200×200 diffractive neurons that modulated the phase of the optical field at their corresponding locations. The transmission modulation of each neuron was determined by the dispersion of the diffractive material and its physical height (see the Methods section), which was optimized using deep learning and error back-propagation[24,26].

A single square pixel with a width of $3\lambda_{max}$ was placed $8.3\lambda_{max}$ away from the last diffractive layer, measuring the intensity of all the 20 pre-determined wavelengths. This can be achieved sequentially by, e.g., wavelength scanning or turning different sources on/off; alternatively, it can run simultaneously by having a spectroscopic detector behind the single-pixel aperture. The measured single-pixel power spectrum $s = [s_0, s_1, ... s_{19}]$ for these 20 wavelengths was paired in groups of two to differentially represent each spectral class score[25,47]. The first 10 spectral measurements at the single-pixel detector ($s_0$, $s_1$, ... , $s_9$) were virtually assigned to be the positive signals ($s_{0,+}$, $s_{1,+}$, ... , $s_{9,+}$) and the subsequent 10 spectral measurements ($s_{10}$, $s_{11}$, ... , $s_{19}$) were virtually assigned to be the negative signals ($s_{0,-}$, $s_{1,-}$, ... , $s_{9,-}$). Based on this, the differential spectral class score $\Delta s_c$ for a given data class $c$ can be defined as:

$$\Delta s_c = \frac{1}{T} \cdot \frac{s_{c,+} - s_{c,-}}{s_{c,+} + s_{c,-}} \qquad (1),$$

where $T$ is a fixed parameter, set as 0.1. A *max(.)* operation on $\Delta s_c$ infers the final classification decision for the input object (see the Methods section for details).

The deep learning-based training process enables the diffractive networks to classify input objects through random unknown diffusers. Each training iteration starts with randomly selecting **B**=4 digits from the MNIST training dataset (containing 50,000 handwritten digits) to form a training batch. The optical fields of the objects in each batch were independently propagated (at the 20 wavelengths of interest) to a phase-only random diffuser. The distorted fields were further propagated, modulated by three successive diffractive layers, and reached the single-pixel detector at the output. A softmax cross-entropy loss[26] was calculated using the differential spectral class scores ($\Delta s_c$) and the ground truth class labels to update the neurons' height profiles through error back-propagation, which concluded one training batch. A training epoch finished when all the 50,000 training handwritten digits were used, i.e., after 12,500 batches. Within each epoch, *n* different random phase diffusers were used to ensure generalization to classify new test objects through unseen, new diffusers (Fig. 2(b)). Therefore, the random diffuser in the forward model was regularly updated after every ~12,500/*n* training batches within each epoch. Each diffractive network was trained for 100 epochs; during its training, each diffractive network 'saw' *N*=100*n* different random diffusers (referred to as *known* diffusers).

The trained single-pixel diffractive network is able to blindly classify handwritten objects through not only the diffusers used during the training (i.e., the known diffusers) but also new, random phase diffusers that were never seen by the network (see Fig. 3). For example, the diffractive network trained with *n*=80 and *N*=8000 different random diffusers achieved an average blind testing accuracy of 92.30±1.67% classifying handwritten test digits through the same 80 known diffusers used in the *last training epoch;* the same single-pixel broadband diffractive network achieved an average blind testing accuracy of 88.53±4.02% classifying handwritten test digits through $N_t$=80 new random phase diffusers, never used in the training phase. This reduction in the handwritten digit classification accuracy of the diffractive network through new random diffusers, compared to the known diffusers used in the last epoch, indicates that the network overfitted to, or 'memorized', the random diffusers used in the last epoch. To shed more light on this, we further divided the known random diffusers into two categories: the *memorized* diffusers are the *n* random diffusers used in the last epoch of the training (epoch 100), and the *forgotten* random diffusers are those used in the training epochs 1-99 (except the last epoch). In terms of the input object classification performance, the single-pixel diffractive network treats the earlier training diffusers the same as the new ones: the single-pixel broadband diffractive network achieved a blind testing classification accuracy of $Acc_m$ = 92.30±1.67% through 80 memorized random diffusers used in epoch 100, $Acc_f$=88.36±4.34% through 7920 forgotten diffusers used in epochs 1-99, and $Acc_{new}$=88.53±4.02% through 80 new random diffusers never used in the training phase. In addition, the classification accuracy of the same diffractive network when the random diffusers were removed was calculated to be $Acc_0$ = 95.78%.

From these analyses we conclude that $Acc_0 > Acc_m > Acc_f \approx Acc_{new}$, which indicates that (1) the single-pixel broadband diffractive network trained with random phase diffusers can classify input objects more accurately when there are no diffusers in the testing phase, showing that it converged to a decent single-pixel image classifier; (2) it partially memorized the random diffusers of the last epoch and performed better all-optical image classification through these memorized diffusers compared to the forgotten diffusers of the previous epochs; and (3) it performed at a similar level of classification accuracy for new random phase diffusers when compared to the forgotten diffusers since $Acc_f \approx Acc_{new}$. This brings more meaning to the term "forgotten diffuser" as it is statistically equivalent to a new random diffuser from the perspective of the broadband diffractive network's image classification performance. Figure 4 supports the same conclusions, reporting the confusion matrices for diffractive single-pixel image classification without a diffuser as well as through the memorized, forgotten, and new random unknown diffusers.

The single-pixel broadband diffractive network's generalization capability, or its resilience to random new diffusers' distortions, strongly correlates with the number of diffusers used in its training. To better highlight this feature, we further trained three additional single-pixel diffractive networks with **n**=10, 20 and 40 random diffusers in each epoch (i.e., **N**=1000, 2000 and 4000, respectively), and the resulting $Acc_m$, $Acc_f$, $Acc_{new}$ and $Acc_0$ values are compared in Fig. 5. With **n**=10 (**N**=1000), the diffractive network obtained a strong memory for classification through diffusers, yielding $Acc_m$=93.89±2.19%. However, the generalization capability was consequently limited, with $Acc_{new}$ =78.17 ± 10.13%. An improved generalization over unknown random diffusers can be obtained when the network is trained with an increased number of diffusers in each epoch. For example, $Acc_{new}$ values increased to 81.39±9.03%, 85.67±6.58% and 88.53±4.02% when **n** = 20, 40 and 80, respectively. At the same time, the capability to classify objects without diffusers remained largely unchanged, with $Acc_0$ being 96.62% for **n**=10 and 95.78% for **n**=80. These results indicate that the diffractive network trained with a larger **n** learned the image classification task through random new diffusers better, converging to a state where more of the diffractive features were utilized to accommodate for the existence of a random phase diffuser for correct image classification through a single-pixel output detector.

**Image classification through random unknown diffusers with different correlation lengths**

To demonstrate the applicability of the presented framework under different levels of image distortion, we further trained five new diffractive networks with different correlation lengths, i.e., we used an $L_{train}$ of 3.2$\lambda_{max}$, 10.9$\lambda_{max}$, 15.1$\lambda_{max}$, 33.8$\lambda_{max}$ and 62.3$\lambda_{max}$. All these single-pixel broadband diffractive networks were trained following the same workflow as depicted in Fig.2, only changing the random phase diffusers to create different levels of image distortions. Each one of the trained diffractive networks was separately tested with $N_t$=80 random unknown diffusers with $L_{test}=L_{train}$ (see Fig. 6). Our results indicate that the single-pixel image classification networks that were trained and tested with random phase diffusers with larger correlation lengths achieved better classification accuracies, as shown in Fig. 6a. This improvement is largely owing to the reduced distortion generated by diffusers with a larger correlation length (see Fig. 6b).

For each one of the diffractive networks shown in Fig. 6a, the image classification accuracies through random diffusers once again confirmed that $Acc_m > Acc_f$ and $Acc_f \approx Acc_{new}$, which were all lower than $Acc_0$. In fact, $Acc_0$ was also lower than the classification accuracy of a single-pixel broadband diffractive network that was trained and tested *without* any diffusers, which scored a blind testing accuracy of 98.43% in classifying distortion-free handwritten digits (see the dashed line in Fig. 6a).

It is also worth noting that $Acc_0$ experienced a relatively steep increase with larger correlation lengths. For example, the single-pixel broadband diffractive network designed for classifying input objects through $L_{train} = 3.2\, \lambda_{max}$ random diffusers achieved a blind testing accuracy of $Acc_0$=71.11%, which drastically improved to 92.71% for $L_{train}$=10.9$\lambda_{max}$ and further increased to 97.67% for $L_{train}$=62.3$\lambda_{max}$. This performance increase indicates that the single-pixel broadband diffractive networks trained with random phase diffusers present a trade-off between their image distortion-resilience and diffuser-free image classification. To classify objects through random diffusers with smaller correlation lengths, the diffractive networks optimized most of their diffractive neurons for rejecting the distortions generated by random diffusers, intuitively resulting in a limited number of diffractive units optimized for the all-optical image classification task. With an increased correlation length, however, random diffusers create less distortions to the input optical fields, giving the diffractive networks more degrees of freedom to optimize their diffractive neurons for enhancing their diffuser-free image classification performance, i.e., $Acc_0$. That is why, $Acc_0$ increased to 97.67% and 92.71% from 71.11% when $L_{train}$ increased to $62.3\lambda_{max}$ and $10.9\lambda_{max}$ from $3.2\lambda_{max}$, respectively (Fig. 6a).

## Discussion

Apart from the deep learning-based training strategies we employed, two design features played an important role in achieving high object classification accuracies through random, unknown diffusers: (1) the differential spectral encoding scheme and (2) the use of broadband illumination. Optoelectronic detectors can only detect non-negative optical intensity information, limiting the range of realizable output values of a diffractive network. The differential encoding scheme has been originally proposed to mitigate this constraint by virtually assigning a negative sign to some of the output detectors[25], which was beneficial for various applications of diffractive networks[27,46,49]. To further reveal the importance of our differential spectral detection scheme (Eq. 1) used for image classification through random diffusers, we trained another diffractive network without any differential encoding using the same physical configuration: the size of the diffractive layers, their spatial arrangement, and the training strategy were kept the same as described in Fig. 2, except that we reduced the number of wavelengths to 10 and used their intensity to directly encode the class labels as opposed to the differential scheme used in Eq. 1. This converged diffractive network scored an accuracy 75.15±8.83% in classifying handwritten digits through new random diffusers, which is on average 13.38% lower than the differentially encoded single-pixel diffractive network's performance.

In addition to the differential spectral class encoding strategy (Eq. 1) that we employed, using broadband illumination also provides an additional computational capacity to accurately classify objects through unknown random diffusers. To shed more light on this, we designed another diffractive network utilizing coherent illumination at a *single* wavelength ($\lambda$) to classify handwritten digits through unknown random diffusers following the same procedures as depicted in Fig. 2. Instead of using the single-pixel spectrum detection approach, for this monochrome diffractive network we utilized an array containing 20 optoelectronic detectors spatially distributed on the output plane (see Supplementary Fig. S1 for details). A differential detection scheme was applied to these 20 output detectors to retrieve the differential class scores and the corresponding class prediction[25]. After its training, this differential diffractive network that uses a single wavelength achieved a blind testing accuracy of 78.32±1.95% for classifying handwritten digits through unknown random diffusers with the same correlation length ($25\lambda$); this classification performance is on average 10.21% lower than our broadband single-pixel diffractive network design reported earlier.

These analyses indicate that broadband illumination is the key to a competitive classification accuracy through unknown random diffusers using a diffractive network despite the fact that (1) a single output pixel was used, and (2) the modulation function of a diffractive neuron at different wavelengths are tightly coupled to each other through the material dispersion. For the classification of handwritten objects through random unknown diffusers, using the same number of diffractive neurons, training data and epochs, the broadband single-pixel diffractive network clearly surpassed the performance of the monochrome diffractive network that used 20 output detectors in the differential configuration, proving the improved information processing capacity of a wavelength-multiplexed diffractive network design.

In conclusion, we presented an all-optical processor to classify unknown objects through random, unknown diffusers using a broadband single-pixel diffractive network. Designed to classify handwritten digits through random unknown diffusers, the single-pixel broadband diffractive network memorized the diffusers used in the last training epoch, scoring $Acc_m$=92.30% image classification accuracy when the random diffusers from the last training epoch are used. The same single-pixel broadband diffractive network can also classify blind objects through unknown new diffusers never used in training, achieving an average accuracy of $Acc_{new}$=88.53%. This diffractive framework was also applied to classify objects through random phase diffusers with various correlation lengths, showing an improved classification accuracy when random diffusers with larger correlation lengths were used since the input images were less distorted. The image classification through random diffusers does not require any external computing power except for the illumination source, presenting a time- and energy-efficient solution. The teachings of this all-optical processor will find unique applications in many fields, such as surveillance cameras, security, biomedical imaging and autonomous driving.

## Materials and Methods

**Model of a broadband single-pixel diffractive network**

Broadband illumination was used for object classification through unknown diffusers. The diffractive layers were modeled as thin optical modulation elements, where the $i^{th}$ neuron on the $l^{th}$ layer at a spatial location $(x_i, y_i, z_i)$ represents a wavelength ($\lambda$) dependent complex-valued transmission coefficient, $t^l$, given by:

$$t^l(x_i, y_i, z_i, \lambda) = a^l(x_i, y_i, z_i, \lambda)\exp\left(j\phi^l(x_i, y_i, z_i, \lambda)\right) \quad (2).$$

In this work, we assumed $a^l(x_i, y_i, z_i, \lambda) = 1$. The phase modulation $\phi^l(x_i, y_i, z_i, \lambda)$ can be written as a function of the thickness of each diffractive neuron $h_i^l$ and the incident wavelength $\lambda$:

$$\phi^l(x_i, y_i, z_i, \lambda) = (n(\lambda) - n_{air})\frac{2\pi h_i^l}{\lambda} \quad (3),$$

where $n(\lambda)$ is the refractive index of the diffractive material. In this work, the height of each neuron was defined as:

$$h_i^l = \frac{h_{\max}}{2} \cdot \left(\sin(h_p) + 1\right) + h_{base} \quad (4),$$

where $h_p$ is the latent variable that was optimized during the data-driven training procedure. The ultimate height of each diffractive neuron $h_i^l$ was constrained by setting $h_{max}=0.83\lambda_{max}$, with a fixed base height $h_{base}=0.42\lambda_{max}$.

The diffractive layers were optically connected to each other by diffracted light propagation in free-space, which was modeled through the Rayleigh-Sommerfeld diffraction equation[24,46]. Each neuron $(x_i, y_i, z_i)$ on $l^{th}$ layer can be viewed as a secondary wave source, generating a complex-valued field $w_i^l(x, y, z, \lambda)$ at a spatial location of $(x, y, z)$, which can be formulated as:

$$w_i^l(x, y, z, \lambda) = \frac{z - z_i}{r^2}\left(\frac{1}{2\pi r} + \frac{1}{j\lambda}\right)\exp\left(\frac{j2\pi r}{\lambda}\right) \quad (5),$$

where $r = \sqrt{(x - x_i)^2 + (y - y_i)^2 + (z - z_i)^2}$ and $j = \sqrt{-1}$. For the $l^{th}$ layer ($l \geq 1$), the modulated optical field $u^l$ at location $(x_i, y_i, z_i)$ is given by:

$$u^l(x_i, y_i, z_i, \lambda) = t^l(x_i, y_i, z_i, \lambda) \cdot \sum_{k \in I} u^{l-1}(x_k, y_k, z_k, \lambda) \cdot w_k^{l-1}(x_i, y_i, z_i, \lambda) \quad (6),$$

where $I$ denotes all the diffractive neurons on the previous (i.e., $l - 1^{th}$) diffractive layer. In case of $l = 1$, $u^0(x_k, y_k, z_k, \lambda)$ denotes the optical field right after the random diffuser, which can be formulated as:

$$u^0(x_i, y_i, z_i, \lambda) = t^D(x_i, y_i, z_i, \lambda) \cdot \sum_{k \in I} o(x_k, y_k, z_k, \lambda) \cdot w_k^{OD}(x_i, y_i, z_i, \lambda) \quad (7),$$

where $o(x_k, y_k, z_k, \lambda)$ is the transmission function of a pixel at the input object plane and $w_k^{OD}$ denotes the free-space propagation from the object plane to the diffuser plane. $t^D(x_i, y_i, z_i, \lambda)$ is the modulation generated by a random phase diffuser, which can be calculated using its height map $h_D$ and Eq (3). The height map of each random phase diffuser was defined as:

$$h_D(x, y) = rem\left(W(x, y) * K(\sigma) + h_{base}, \frac{\lambda_{max}}{n(\lambda_{max}) - n_{air}}\right). \quad (8)$$

where $rem(.)$ denotes the remainder after division. $W(x, y)$ is a random height matrix that follows a normal distribution with a mean of $\mu$ and a standard deviation of $\sigma_0$, i.e.

$$W(x, y) \sim \mathcal{N}(\mu, \sigma_0). \quad (9)$$

$K(\sigma)$ is a Gaussian smoothing kernel with zero mean and standard deviation $\sigma$. '$*$' denotes the 2D convolution operation. The correlation length ($L$) of a random diffuser was calculated using the 2D auto-correlation function ($R_d$) of its height profile $h_D(x, y)$, based on the following equation:

$$R_d(x, y) = \exp(-\pi(x^2 + y^2)/L^2) \quad (10)$$

A single-pixel detector was placed on the optical axis at the end of the diffractive network, after the $L^{th}$ layer, which measured the intensity at each encoding wavelength within a square aperture of $3\lambda_{max}$ by $3\lambda_{max}$ The single pixel spectral measurement $s_p$ at a wavelength of $\lambda_p$ can be formulated as:

$$s_p = \left|\sum_{k \in I} u^L(x_k, y_k, z_k, \lambda_p) \cdot w_k^L(x_i, y_i, z_i, \lambda_p)\right|^2 \quad (11).$$

The differential spectral class scores were calculated following Eq. (1) and the diffractive networks were trained to optimize the classification accuracy using a softmax-cross-entropy (SCE) loss:

$$\mathcal{L}_I = -\sum_{c=0}^{9} g_c \cdot \log\left(\frac{\exp(\Delta s_c)}{\sum_{c'=0}^{9} \exp(\Delta s_{c'})}\right) \quad (12),$$

where $\Delta s_c$ denotes the spectral class score for the $c^{\text{th}}$ class, and $g_c$ denotes the $c^{\text{th}}$ entry of the ground truth label vector.

**Digital implementation**

The diffractive neural networks presented here contained 200×200 diffractive neurons on each layer with a pixel size (pitch) of $0.25\lambda_{max}$. During the training, each handwritten digit of the MNIST dataset was first upscaled from 28×28 pixels to 70×70 pixels using bilinear interpolation, and then padded with zeros to cover 200×200 pixels. The broadband illumination was digitally modeled as multiple independently propagating monochrome plane waves; we used $\lambda_{min} = 0.6$ mm and $\lambda_{max} = 1.2$ mm based on the THz part of the spectrum. The propagation and wave modulation on each spectral channel were separately computed. Four different randomly selected MNIST images formed a training batch, providing amplitude-only modulation to the input broadband light. Each input object batch was propagated and disturbed by one randomly selected diffuser. The four distorted broadband fields were separately propagated through the diffractive network, and the loss value (Eq. (12)) was calculated accordingly. The resulting loss was back-propagated, and the pixel height values were updated using the Adam optimizer[50] with a learning rate of $1 \times 10^{-3}$. Our models were trained using Python (v3.7.3) and PyTorch (v1.11) for 100 epochs, which took 5 hours to complete. A desktop computer with a GeForce RTX 3090 graphical processing unit (GPU, Nvidia

Inc.), an Intel® Core ™ i9-7900X central processing unit (CPU, Intel Inc.) and 64 GB of RAM was used.

# Figures and Figure Captions

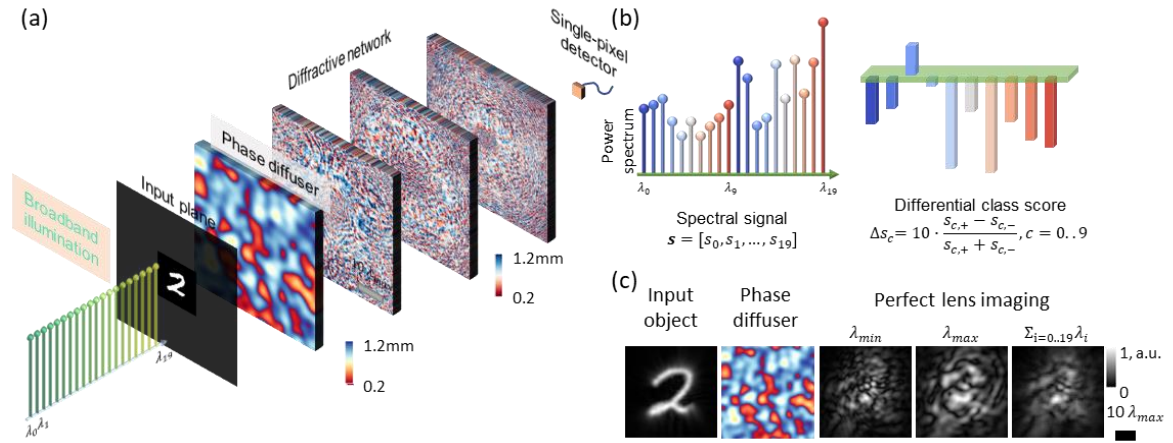

**Figure 1**. A single-pixel broadband diffractive neural network classifies handwritten digits through unknown random diffusers. (a) The schematic drawing of a broadband single-pixel diffractive network mapping the spatial information of an input handwritten digit behind an unknown diffuser into the power spectrum at the output pixel aperture. (b) Differential spectral class score encoding scheme using 20 illumination wavelengths. (c) Visualization of the level of image distortions through a random phase diffuser with a correlation length of $25\lambda_{max}$. The distortions induced by the random diffuser are severe at all the utilized wavelengths between $\lambda_{min}$ and $\lambda_{max}$.

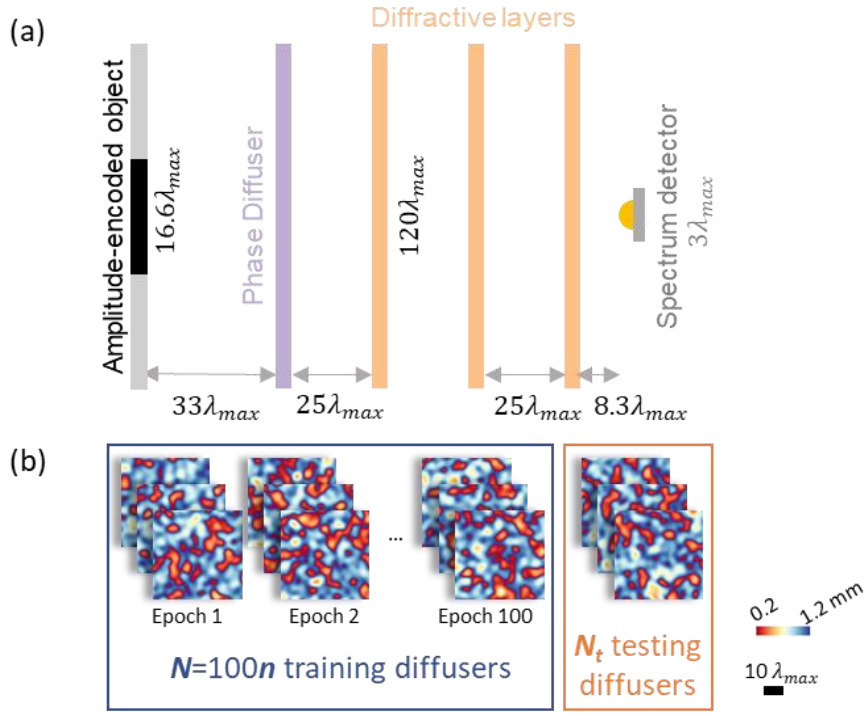

**Figure 2**. Training of a single-pixel broadband diffractive network. (a) The physical design of the diffractive network. (b) During each training epoch, *n* different random diffusers with the same correlation length were used to generate unique distortions to the input objects; *N* = 100*n* training random diffusers are used across 100 epochs. $N_t$ = 80 new random diffusers (never used in training) were used to verify the network's generalization performance after the training.

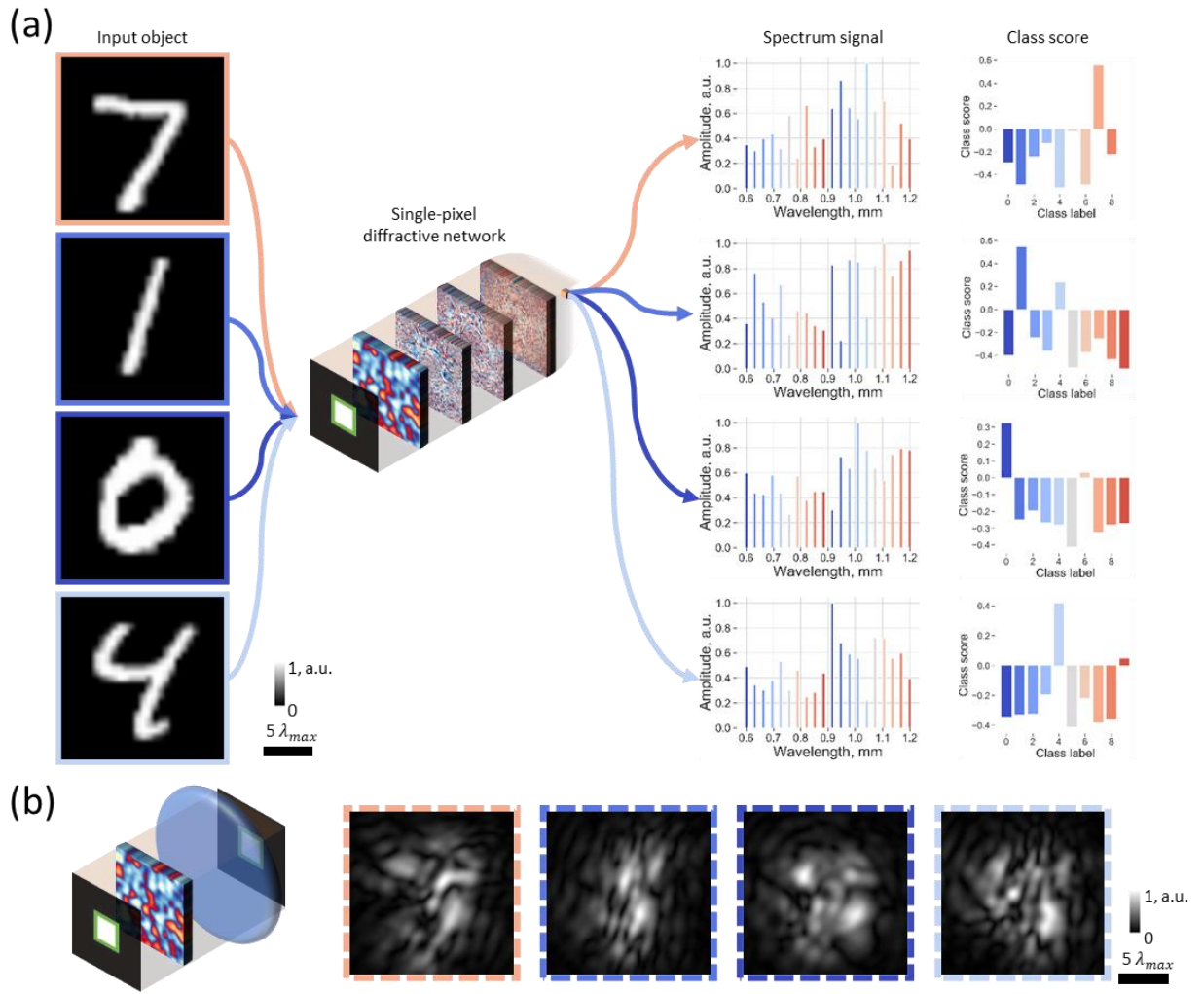

**Figure 3**. (a) Example results for classifying handwritten digits through a new random diffuser using the single-pixel broadband diffractive network. (b) Comparison images, for each test object of (a), seen through the same random diffuser using a diffraction-limited lens at $\lambda_{max}$ illumination (0.8 NA). The spatial information of the input objects is severely distorted, making then unrecognizable.

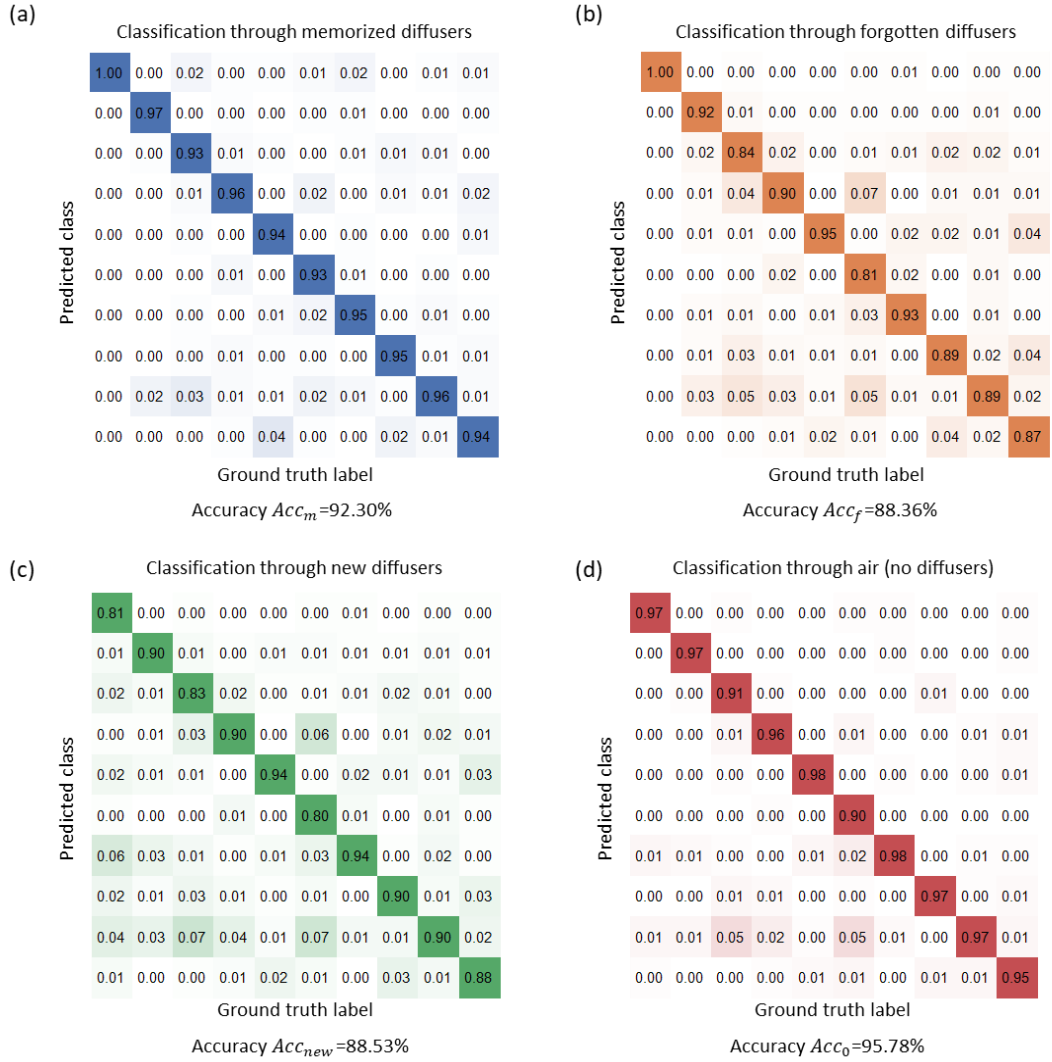

**Figure 4**. Confusion matrices of the single-pixel broadband diffractive network trained with ***n***=80 (***N***=8000) random diffusers classifying handwritten test digits through (a) 80 memorized random diffusers used in the last training epoch, (b) 7920 forgotten random diffusers used in training epochs 1-99, and (c) 80 new random diffusers that were never used in the training of the diffractive network. (d) The confusion matrix of the same single-pixel broadband diffractive network, classifying distortion-free objects with no diffusers present. Notice that $Acc_0 > Acc_m > Acc_f$ and $Acc_f \approx Acc_{new}$. Since $Acc_f \approx Acc_{new}$, we conclude that the diffractive network "forgets" the random training diffusers used in epochs 1-99 and statistically treats them the same as a fresh random diffuser never used during the training.

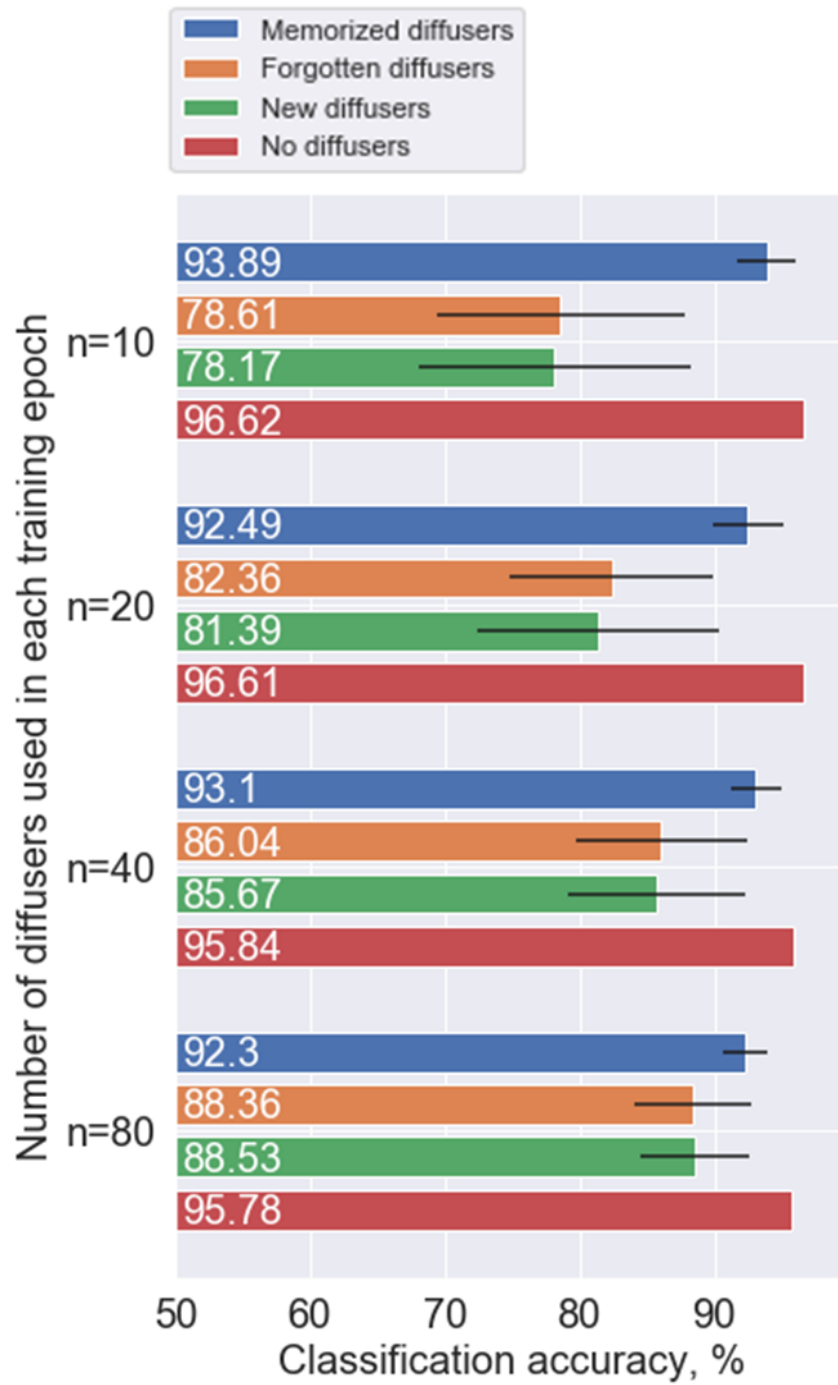

**Figure 5**. Training with additional number of random diffusers in each epoch improved the network's generalization for classifying handwritten digits behind unknown random diffusers using a single output pixel.

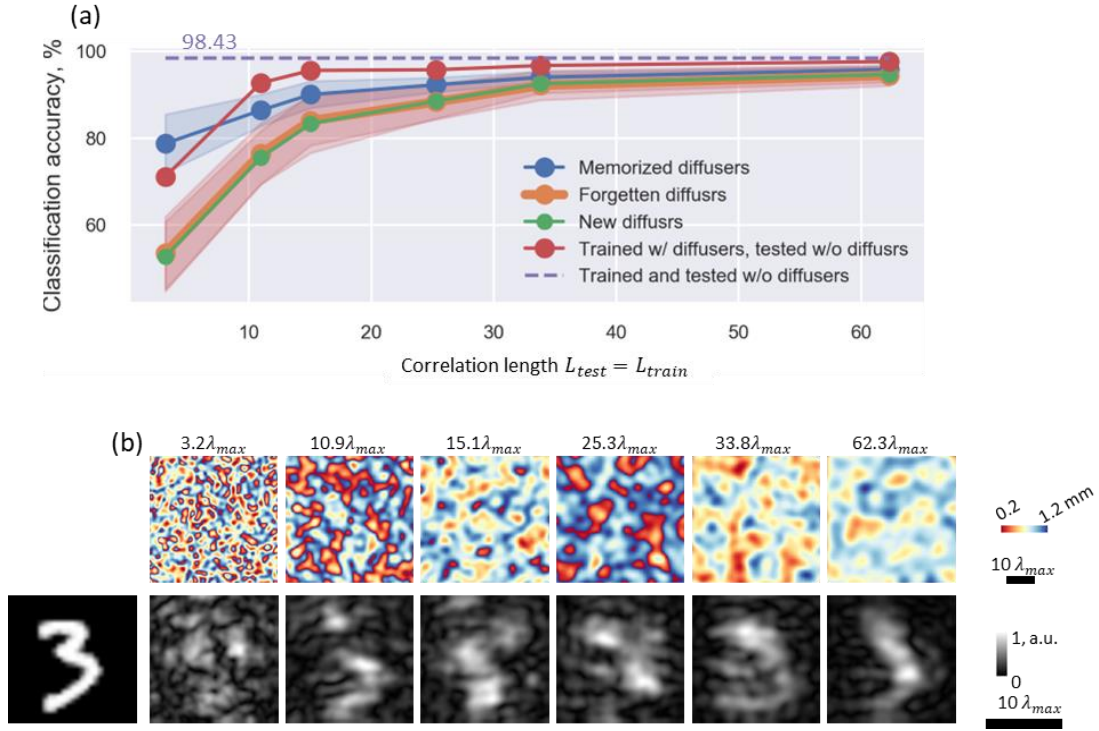

**Figure 6.** Single-pixel broadband diffractive networks trained to classify handwritten digits through random diffusers with different correlation lengths ($L_{train}$). (a) The classification accuracy of single-pixel broadband diffractive networks trained with different $L_{train}$, classifying handwritten test digits through memorized, forgotten, new and no diffusers. The dashed purple line indicates the classification accuracy without any diffusers being present using a broadband diffractive network trained *without* any diffusers. (b) Comparison images of a test object seen through random diffusers with different correlation lengths (from $3.2\lambda_{max}$ to $62.3\lambda_{max}$) using a diffraction-limited lens at $\lambda_{max}$ illumination. The spatial information of the input object is severely distorted in each case, making the object unrecognizable.